\documentclass[a4paper,10pt]{article}
\usepackage{algorithmic,algorithm}
\usepackage{color,colortbl,float,subfig,multicol,textcomp}
\usepackage[colorlinks,citecolor=blue,linkcolor=black]{hyperref}
\usepackage{amsmath,amsfonts,amssymb,graphicx,latexsym}
\usepackage[a4paper,includeheadfoot,pdftex]{geometry}
\usepackage{booktabs}
\usepackage[braket]{qcircuit}
\usepackage{multirow,scalefnt}

\usepackage{calc,tikz}
\usetikzlibrary{decorations.pathreplacing,decorations.markings,calc}

\newtheorem{example}{Example}
\newcommand{\etal}{\emph{et al. }}
\title{Quantum associative memory with linear and non-linear algorithms for the 
diagnosis of some tropical diseases}

\author{J-P. TCHAPET NJAFA, S.G. NANA ENGO \\
\small{Laboratory of Photonics, Department of Physics,} \\\small{University of
Ngaoundere, PO. BOX 454 Ngaoundere, Cameroon}
}

\begin{document}
\maketitle
\begin{abstract}

This paper presents the QAMDiagnos, a model of Quantum Associative Memory (QAM)
that can be a helpful tool for medical staff without experience or laboratory
facilities, for the diagnosis of four tropical diseases (malaria, typhoid fever,
yellow fever and dengue) which have several similar signs and symptoms. The
memory can distinguish a single infection from a polyinfection. Our model is a
combination of the improved versions of the original linear quantum retrieving
algorithm proposed by Ventura and the non-linear quantum search algorithm of
Abrams and Lloyd. From the given simulation results, it appears that the
efficiency of recognition is good when particular signs and symptoms of a
disease are inserted given that the linear algorithm is the main algorithm. The
non-linear algorithm helps confirm or correct the diagnosis or give some advice
to the medical staff for the treatment. So, our QAMDiagnos that has a friendly
graphical user interface for desktop and smart-phone is a sensitive and a
low-cost diagnostic tool that enables rapid and accurate diagnosis of four
tropical diseases.
\end{abstract}

\section{Introduction}
Diagnosis is the identification of a situation, a difficulty or a phenomenon by
interpreting external signs or lesions. In the medical field, it consists of
identifying abnormal condition that afflicts a specific patient, based on
manifested clinical data. If the final diagnosis matches a disease that afflicts
a patient, the diagnostic process is correct; otherwise, a misdiagnosis occurs.
Medical diagnosis can also be defined as the process that allows a physician,
through information collected from a history and physical examination of a
person, to make prediction about features of clinical situations and determine
appropriate course of action. It implements a complex decision process that
involves a lot of vagueness and uncertainty management, especially when a
disease has multiple signs and symptoms or non-specific signs and symptoms.

From the point of view of statistics, the diagnostic procedure involves
classification tests. That is, the task is made on the basis of measured
features to assign a patient to one of a small set of classes \cite{Dybowski}.
Then, Artificial Neural Networks (ANN) provide powerful tools that help
physicians avoid misdiagnosis by analyzing, modelling and making sense of
complex clinical data across a broad-range of medical applications. As ANNs have
the ability of prediction, parallel operation and adaptability, they have been
widely used as computer-assisted tools among many techniques about automatic
disease diagnosis proposed in the literature \cite{Filippo, AlkimGK12}. ANNs 
have
been used for example for the diagnosis of colorectal cancer, multiple sclerosis
lesions, colon cancer, tuberculosis, pancreatic disease, gynecological diseases
and early diabetes \cite{Filippo}.

Associative Memories (AM) are a class of ANNs specialised in pattern recognition
that have drawn the attention of major research groups in the world due to a
number of properties such as a rapidity, a compute efficient best-match and an
intrinsic noise tolerance. Aldape-P\'erez \etal gave in the ref. \cite{Aldape} a
good summary of the AM evolution, from the first model, the Lernmatrix developed
by Karl Steinbuch in 1961, to the recent model proposed between 1982-1984 by
Hopfield \cite{Hopfield82, Hopfield84}. The latter is an AM that uses the
Hebbian learning rule and it is able to recall patterns from noisy or partial
representation.

Some computer-assisted tools in the case of tropical diseases, are already built
and some of them use ANNs, but they are mainly specialized for malaria
\cite{Sunny,neetu,Khalda,andrade,Filippo}. It is worth noting that malaria is the
most parasitic disease spread over the world. $40\%$ of the world's population
 are concerned, especially those of tropical regions. In Cameroon, as in the
 most Sub-Saharan countries, it is a public health problem because the whole
 population is exposed to the disease. To diagnose malaria, the World Health
 Organisation (WHO) \cite{who} recommends the use of rapid diagnostic tests.
 However these tools need some conservation facilities that are difficult to
 find in rural and semi-urban regions in developing countries. So, the most
 widely used technique for determining the development stage of malaria is
 through visual microscopical evaluation of Giemsa stained blood smears.
 However, this is a routine and time-consuming task and it requires a
 well-trained operator. In addition, there is frequently a misdiagnosis because
 of confusion between signs and symptoms of malaria and that of other tropical
 diseases such as typhoid fever, yellow fever and dengue, or inexperience of
 medical staff. As malaria, these tropical diseases are also life-threatening.

In order to provide a rapid and accurate tool for the diagnosis of the above
mentioned four tropical diseases (malaria, typhoid fever, yellow fever and
dengue), we propose the framwork QAMDiagnos, Quantum Associative Memory for the
Diagnosis, associated with a friendly multi-platform graphical user interface
(Android, Linux, MS Windows). It is a combination of improved versions of the
original linear quantum retrieving algorithm proposed by Ventura
\cite{tchapet2012} and the non-linear quantum search algorithm of Abrams and
Lloyd \cite{tchapet2016}. The aim of QAMDiagnos is to (i) act as an advisory
tool to inexperienced medical staff, especially senior nurses in rural health
centres having no or a limited number of physicians; (ii) act as a
decision-support tool for medical diagnosis for physicians in under-staffed
health centres; (iii) provide an alternative way to reach a reasonable tentative
diagnosis, and hence early commencement of clinical management of patients in
the absence of laboratory facilities in many rural and semi-urban health
centres; (iv) facilitate the treatment and prevent potential pandemics given the
fact that today the increase of international air travel/traffic, tourism to
tropical regions and human migration led to a rising incidence of tropical
diseases (since vaccines are unavailable for most major tropical infections).
Our framework, which is more robust than that of Agarkar and Ghathol
\cite{Agarkar} that uses the FFANN for the diagnosis of malaria, typhoid fever
and dengue, can be extended to a wide range of tropical diseases.

The paper is structured as it follows: Section \ref{sec:qam} provides a
description of foundations of our model of Quantum Associative Memory. In
Section \ref{sec:dtb} a brief description of each disease is given. Signs and
symptoms of diseases are given in Appendix \ref{sec:signs}. Section
\ref{sec:Sim} is devoted to simulations results and discussion whereas in
Section \ref{sec:over} we explain the use of the multi-platform friendly graphical user
interface (GUI) of QAMDiagnos, designed for the medical staff. Finally, we
conclude with an outlook of possible future improvements.

\section{Description of Quantum Associative Memory}\label{sec:qam}

Associative Memories are a class of Artificial Neural Networks that can memorise
information, field of knowledge or patterns and can retrieve that from partial 
or noisy data. Quantum Associative Memories (QAM) combine neurocomputing with
quantum computations. Therefore, QAM models share main features both of quantum 
information theory and Associative Memory. 

\subsection{Few basic concepts of quantum information theory}

\subsubsection{Quantum bit}
Quantum information use a \emph{quantum bit} or a \emph{qubit} instead 
of an ordinary bit as the fundamental unit of information. The qubit can be an 
atom, a molecule or a photon that can be in a superposed state (energy or spin 
for example). As the classical bit, the qubit can take two particular states 
noted $\ket{0}$ and $\ket{1}$ that are the basis states of a Hilbert space
$\mathcal{H}$ of $2$ dimensions. The fundamental difference between the qubit
and the classical bit is that the qubit can take simultaneously both values. 
So, the state of the qubit can be represented by the following \emph{superposed 
state}:
\begin{subequations}
\begin{equation}
\ket{\psi} =\alpha\ket{0}+\beta\ket{1},
\end{equation}
with at any time
\begin{equation}
|\alpha|^{2}+|\beta|^{2}=1,\text{\ \ \ \ }\alpha,\beta\in
\mathbb{C}.
\end{equation}
\end{subequations}
$|\alpha|^2$ and $|\beta|^2$ are the probabilities of the qubit to be found in
the  states $\ket{0}$ and $\ket{1}$ respectively after a measurement. $(\ket{0},
\ket{1})$ is the most used computational basis that is a pair of orthonormal 
vectors defined as
\begin{equation}
	\ket{0}=\begin{pmatrix}1\\0\end{pmatrix},\text{\ \ \ \ } 
	\ket{1}=\begin{pmatrix}0\\1\end{pmatrix}.
\end{equation}
Therefore, quantum superposition suggest that an $n$-qubits register can exist 
in all its possible $2^n$ states at the same time.

\subsubsection{Quantum entanglement}
The use of only one qubit does not highlight the power of quantum information. A
purely quantum phenomenon that has non classical analogue is a \emph{quantum 
entanglement}. We are talking about the quantum entanglement when at least a
pair of qubits has quantum correlation. It means that any interaction with one
of the qubits affects instantaneously the other despite the separation distance,
highlighting the non-local features of quantum theory. In this case, we should
see the whole qubits as one unique physical system instead of many separated
subsystems. So, no classical consideration can allow to know the state of each
qubit. A well-known entangled states are the following EPR states:
\begin{equation}
  \ket{\varPsi^+}=\frac{1}{\sqrt{2}}\left(\ket{00}+\ket{11}\right),
  \ket{\varPsi^-}=\frac{1}{\sqrt{2}}\left(\ket{00}-\ket{11}\right),
  \ket{\varPhi^+}=\frac{1}{\sqrt{2}}\left(\ket{01}+\ket{10}\right),
  \ket{\varPhi^-}=\frac{1}{\sqrt{2}}\left(\ket{01}-\ket{10}\right),
 \label{eq:epr}
\end{equation}
that cannot be written as the products of independent states of two separate 
subsystems.

\subsubsection{Quantum parallelism and decoherence}
Due to the quantum superposition and the quantum entanglement it is possible to 
perform multiple computations simultaneously or \emph{quantum parallelism} on a 
system. But a measurement on the system destroys the quantum superposition and 
the system collapses to one of its possible states. This phenomenon is the 
\emph{quantum decoherence} that can be seen as an interaction between a qubit 
and its environment. One of the challenges of quantum processing consists of 
increasing the probability to observe a needed state before the quantum 
decoherence occurs.

\subsubsection{Elementary quantum gate}
Quantum computing (that is changing the state of a qubit to another) is carried 
out through \emph{unitary operators} or \emph{quantum logic gates} when they 
are used in quantum circuit. Unlike many classical logic gates, the quantum 
logic gates are reversible, and therefore allow to avoid energy dissipation. 
Let us recall that the operator $\mathtt{U}$ is an unitary operator if
 \begin{equation}
  \mathtt{U}\mathtt{U}^{\dagger}=\mathtt{U}^{\dagger}\mathtt{U}=\mathbb{I},
 \end{equation}
 where $\mathbb{I}$ is the identity operator, and $ \mathtt{U}^{\dagger} $ the 
complex conjugate
 transpose of $\mathtt{U}$. Any unitary operator can be written as 
 \begin{equation}
  \mathtt{U}=\exp{\left(-i\mathtt{G}\right)},
 \end{equation}
where $\mathtt{G}$ is an hermitian operator, that is $\mathtt{G}=\mathtt{G}^\dagger$.
 
 In the present work, the $\mathtt{NOT}$ gate $ \mathtt{X}=\ket{1-x}\bra{x}$ 
and the Walsh-Hadamard gate 
$\mathtt{W}=\frac{1}{\sqrt{2}}\left((-1)^{x}\ket{x}\bra{x}+\ket{1-x}\bra{x}
\right)$ will be the most using single-qubit quantum logic gates:
 \begin{equation}
 \Qcircuit @C=1em @R=.7em {
 	\lstick{\ket{x}} & \gate{\mathtt{X}} & \rstick{\ket{1-x}}\qw
 }
 \text{\hspace*{6em}} \mathtt{X}=\begin{pmatrix} 0 & 1\\1 & 0\end{pmatrix}\\
 \end{equation}
 \begin{equation}
 \Qcircuit @C=1em @R=.7em {
 	\lstick{\ket{x}} & \gate{W} &
 	\rstick{\frac{1}{\sqrt{2}}((-1)^{x}\ket{x}+\ket{1-x})} \qw
 }
 \text{\hspace*{12em}}  \mathtt{W}=\frac{1}{\sqrt{2}}\begin{pmatrix}1 & 1\\1 & 
 -1\end{pmatrix},
 \end{equation}
 where $ x\in\{0,1\} $.
 
 The two-qubit quantum logic gate mostly used in this work will be the
 controlled $\mathtt{NOT}$ gate $ 
 \mathtt{CX}=(\ket{0}\bra{0})\otimes\mathbb{I}+(\ket{1}\bra{1})\otimes\mathtt{X}
 $:
 \begin{equation}
 \Qcircuit @C=1em @R=1.4em {
 	\lstick{\ket{x}}  &  \ctrl{1} &  \rstick{\ket{x}} \qw \\
 	\lstick{\ket{y}}  &  \gate{\mathtt{X}}  & \rstick{\ket{1-y}} \qw 
 }
 \text{\hspace*{6em}} 
 \mathtt{CX}=\begin{pmatrix}\mathbb{I} & \mathbb{O}\\ \mathbb{O} & \mathtt{X} 
 \end{pmatrix}
 =\left(\begin{array}{cc|cc}
 1 & 0 & 0 & 0\\
 0 & 1 & 0 & 0\\\hline
 0 & 0 & 0 & 1\\
 0 & 0 & 1 & 0
 \end{array}\right),
 \end{equation}
 where $x,y\in\{0,1\}$.  The $\mathtt{CX}$ gate acts on two-qubits and  performs
the $ \mathtt{X} $ operation on the second qubit only when the first qubit is
in the state $\ket{1}$, and otherwise leaves it unchanged.
  
 \subsection{Few basic concepts of Associative Memories}
A Neural Network, more properly referred to as an \emph{Artificial Neural
Network} (ANN), is a computing system made up of several important basic
elements, which include the concept of a processing element (neuron), the 
transformation performed by this element (in general, input summation and
non-linear mapping of the result into an output value), the interconnection
structure between neurons, the network dynamics, and the learning rules that
govern the modification of interconnection strengths \cite{ezhov2000}. A major 
dichotomisation of Neural Networks can be realised by considering whether they 
are trained in a supervised or unsupervised manner. An example of the latter is 
the Hopfield model of content-addressable memory, or Associative Memory, using 
the concept of attractor states. This model has an apparent similarity to 
human episodic memory: it can recall patterns after a single exposure using a 
Hebbian learning rule, and it is capable to retrieve the above mentioned 
patterns from partial information, partial and noisy information or noisy 
ones.
 
In short, Associative Memories are Neural Networks centralised around
two algorithms. The first is to memorise information or patterns and known as
\emph{learning algorithm}. The second is for the restitution of learned
information or patterns from partial or noisy data. It is known as
\emph{retrieving algorithm}.

 Related to quantum theory, Associative Memories are called Quantum Associative
Memories (QAM) where the learning and retrieving algorithms are quantum
algorithms. The QAM is one of the most promising approaches to quantum
neurocomputing. The linear part of the QAM design here is an improved version
of the one built by Ventura and Martinez where the stored patterns are
considered as the basis states of the memory quantum state \cite{ventura2000}.

It should be noted that in the classical Hopfield network the existence
of symmetric, Hebbian connections, guarantees the stability of a unique stored
pattern; similarly, in a quantum analogue of the Hopfield network the 
quantum entanglement ensures the integrity of a stored pattern (basis state).

The Tab. \ref{tab:Corr} summarises the analogies used in developing a QAM 
\cite{ezhov2000}.

\begin{table}[thpb]
\centering
\begin{tabular}{p{6 cm}l}\hline
	\textbf{Classical Neural Networks} & \textbf{Quantum Associative 
Memory} \\\hline
	Neuronal state $ x_i\in\{0,1\} $ & Qubit $
	\ket{x}=\alpha\ket{0}+\beta\ket{1}$\\
	Connections $\{w_{ij}\}_{ij=1}^{p-1} $ & Quantum entanglement 
$\ket{x_0x_1\ldots
x_{p-1}}$\\
	Learning rule $\sum_{s=1}^p x_i^s x_j^s$ &  Superposition of entangled 
states  $\sum_{s=1}^p \alpha_s\ket{x_0x_1\ldots x_{p-1}}$\\
	 Winner search $ n=\max_i\arg(f_i) $  &  Unitary transformation $ 
	\mathtt{U}\ket{\psi}=\ket{\psi'}$\\
	Output result $ n $ & Decoherence $ 
	\sum_{s=1}\alpha_s\ket{x^s}\Rightarrow \ket{x^k}$ \\\hline 
\end{tabular}
\caption{Corresponding concepts from the domains of classical Neural Networks 
and Quantum Associative Memory.}
\label{tab:Corr}
\end{table}

It is worth noting that there are some other interesting models of quantum 
inspired neural networks and quantum inspired evolutionary optimization 
algorithms 
\cite{Platel:2009:QEA:1720414.1720417,Schliebs2009623,Schliebs,Kasabov10}.

In Subsections \ref{sub_Sec:lear} and \ref{sub_Sec:retri} we will briefly 
describe the quantum learning and quantum retrieving algorithms used on our QAM 
model. The full details can be found in refs. \cite{tchapet2012} and 
\cite{tchapet2016}.

\subsection{Quantum learning algorithm}\label{sub_Sec:lear}
For our QAM an operator name $\mathtt{BDD}$ is used as the learning algorithm.
The $\mathtt{BDD}$ operator is obtained by using the Binary Superposed Quantum
Decision Diagram (BSQDD) proposed by Rosenbaum \cite{Rosenbaum2010}. Contrarily
to other quantum learning algorithms such as the one of Ventura
\cite{ventura99}, which needs the initial state to be $\ket{00\dots0}$, the
BSQDD is computed by using any basis states $\ket{z}$ of the Hilbert space of
$2^n$ dimensions. The idea behind the BSQDD is to represent a quantum
superposition as a decision diagram where each node corresponds to a gate. The
gate that corresponds to the node on each branch of the BSQDD is controlled by
the path 
used to reach it from the root of the decision diagram. Thereby three steps,
given by Algorithm \ref{alg:learn}, are needed to construct one BSQDD.

\begin{algorithm}[htpb]
\caption{Linear QAM retrieving algorithm with distributed query for diagnosis}
\label{alg:learn}
\begin{algorithmic}[1]
\STATE \textbf{Finding the unsimplified BSQDD by using the Hadamard gates,
	Feymann gates, and inverters}; \COMMENT{The number of nodes of this unsimplified
	BSQDD represents the upper bound on the number of the gates that the quantum
	array generated by the BSQDD needs for being constructed.}
\STATE \textbf{Reducing the BSQDD to obtain the final BSQDD}; \COMMENT{The goal
	is to have the lower bound on the number of quantum gates. To achieve this goal,
	we need to merge some nodes (gates) according to the links that can occur
	between qubits (like control qubit and target qubit). Two rules are used to
	obtain the final BSQDD. The first rule states that in two different branches of
	different nodes that correspond to the same next node, that same nodes merge.
	The second rule states that in different branches of different nodes that
	generate the same branch, that same branches merge.}
\STATE \textbf{Converting the BSQDD to a quantum array that generates the
	desired quantum state}. \COMMENT{The array is obtained by adding the gates that
	correspond to the nodes in each layer of the final BSQDD. The starting point is
	the last layer and we always place the new gates to the right of the previously
	placed gates in the quantum array.}
\end{algorithmic}
\end{algorithm}

\begin{example}
Fig. \ref{fig:learn1} gives the three steps allowing to construct the state 
$\sqrt{\frac{1}{5}}(\ket{000}+\ket{010}+\ket{110}+\ket{001}+\ket{101})$ from the
starting state $ \ket{000}$. The elementary gates used are respectively the 
rotation gates 
\begin{equation}
\mathtt{R}(\theta) =\begin{pmatrix}\sqrt{\frac{3}{5}} & \sqrt{\frac{2}{5}}\\ 
\sqrt{\frac{2}{5}} & -\sqrt{\frac{3}{5}}\end{pmatrix},\, 
\mathtt{R}(\alpha)=\begin{pmatrix}\sqrt{\frac{2}{3}} & \frac{1}{\sqrt{3}}\\ 
\frac{1}{\sqrt{3}} & -\sqrt{\frac{2}{3}} \end{pmatrix}, 
\end{equation}
 the Hadamard gate $\mathtt{W}$, and the $\mathtt{NOT}$ gate $\mathtt{X}$. Then 
$\ket{\psi_3}=\sqrt{\frac{3}{5}}\ket{000}+\sqrt{\frac{2}{5}}\ket{100}$ and 
$\ket{\psi_2}=\sqrt{\frac{2}{5}}\ket{000}+\frac{1}{\sqrt{5}}\ket{010}+\frac{1} 
{\sqrt{5}}\ket{100}+\frac{1}{\sqrt{5}}\ket{110}$.
 
\begin{figure}[!htbp]
\centering
\leavevmode
 \subfloat[Unsimplified BSQDD]{
 {\scalefont{0.85}

\begin{tikzpicture}[y=0.80pt, x=0.80pt, yscale=-1.000000, xscale=1.000000, inner 
sep=0pt, outer sep=0pt]
\path[draw=black,line join=miter,line cap=butt,miter limit=4.00,line
  width=0.616pt] (175.5000,96.0122) circle (0.4657cm);

\path[cm={{0.55,0.0,0.0,0.55,(175.5000,96.0122)}},fill=black] (0.0000,0.0000)
  node[] (flowRoot3757-45-3-2-8-6-5) {$\mathtt{R(\theta)}$};

\path[draw=black,line join=miter,line cap=butt,miter limit=4.00,line
  width=0.660pt] (175.5000,68.5122) -- (175.5000,79.5122);

\path[draw=black,line join=miter,line cap=butt,miter limit=4.00,line
  width=0.660pt] (158.2266,96.0122) -- (114.2266,129.0122);

\path[draw=black,line join=miter,line cap=butt,miter limit=4.00,line
  width=0.660pt] (192.0000,96.0122) -- (263.5000,129.0122);

\path[draw=black,line join=miter,line cap=butt,miter limit=4.00,line
  width=0.616pt] (113.0504,146.3618) circle (0.4657cm);

\path[cm={{0.55,0.0,0.0,0.55,(113.0504,146.3618)}},fill=black] (0.0000,0.0000)
  node[] (flowRoot3757-45-3-2-8-6-5-4) {$\mathtt{R(\alpha)}$};

\path[draw=black,line join=miter,line cap=butt,miter limit=4.00,line
  width=0.616pt] (264.0008,145.8609) circle (0.4657cm);

\path[cm={{0.55,0.0,0.0,0.55,(264.0008,145.8609)}},fill=black] (0.0000,0.0000)
  node[] (flowRoot3757-45-3-2-8-6-5-4-2) {$\mathtt{W}$};

\path[draw=black,line join=miter,line cap=butt,miter limit=4.00,line
  width=0.660pt] (96.2779,145.1317) -- (52.2779,178.1317);

\path[draw=black,line join=miter,line cap=butt,miter limit=4.00,line
  width=0.660pt] (130.0513,145.2299) -- (153.5000,178.5122);

\path[draw=black,line join=miter,line cap=butt,miter limit=4.00,line
  width=0.660pt] (247.5231,145.1440) -- (226.1000,178.5122);

\path[draw=black,line join=miter,line cap=butt,miter limit=4.00,line
  width=0.660pt] (281.2965,145.2423) -- (296.5000,178.5122);

\path[draw=black,line join=miter,line cap=butt,miter limit=4.00,line
  width=0.616pt] (51.7008,195.3609) circle (0.4657cm);

\path[cm={{0.55,0.0,0.0,0.55,(51.7008,195.3609)}},fill=black] (0.0000,0.0000)
  node[] (flowRoot3757-45-3-2-8-6-5-4-2-0) {$\mathtt{W}$};

\path[draw=black,line join=miter,line cap=butt,miter limit=4.00,line
  width=0.616pt] (156.2008,194.9837) circle (0.4657cm);

\path[cm={{0.55,0.0,0.0,0.55,(156.2008,194.9837)}},fill=black] (0.0000,0.0000)
  node[] (flowRoot3757-45-3-2-8-6-5-4-2-7) {$\mathbb{I}$};

\path[draw=black,line join=miter,line cap=butt,miter limit=4.00,line
  width=0.616pt] (226.6008,194.9837) circle (0.4657cm);

\path[cm={{0.55,0.0,0.0,0.55,(226.6008,194.9837)}},fill=black] (0.0000,0.0000)
  node[] (flowRoot3757-45-3-2-8-6-5-4-2-2) {$\mathtt{X}$};

\path[draw=black,line join=miter,line cap=butt,miter limit=4.00,line
  width=0.616pt] (297.0008,194.9837) circle (0.4657cm);

\path[cm={{0.55,0.0,0.0,0.55,(297.0008,194.9837)}},fill=black] (0.0000,0.0000)
  node[] (flowRoot3757-45-3-2-8-6-5-4-2-00) {$\mathbb{I}$};

\path[draw=black,line join=miter,line cap=butt,miter limit=4.00,line
  width=0.660pt] (34.8204,195.6998) -- (27.0000,228.0122);

\path[draw=black,line join=miter,line cap=butt,miter limit=4.00,line
  width=0.660pt] (68.0044,195.6016) -- (76.5000,228.0122);

\path[draw=black,line join=miter,line cap=butt,miter limit=4.00,line
  width=0.660pt] (155.7000,211.5122) -- (155.7000,228.0122);

\path[draw=black,line join=miter,line cap=butt,miter limit=4.00,line
  width=0.660pt] (298.7000,211.1350) -- (298.7000,227.6350);

\path[draw=black,line join=miter,line cap=butt,miter limit=4.00,line
  width=0.660pt] (226.1000,211.5122) -- (226.1000,228.0122);

\path[fill=black,line join=miter,line cap=butt,line width=0.800pt]
  (85.5000,115.8622) node[above right] (text6869)
  {$\sqrt{\frac{3}{5}}\ket{000}$};

\path[fill=black,line join=miter,line cap=butt,line width=0.800pt]
  (241.5000,115.8622) node[above right] (text6869-2)
  {$\sqrt{\frac{2}{5}}\ket{100}$};

\path[fill=black,line join=miter,line cap=butt,line width=0.800pt]
  (28.5000,165.8622) node[above right] (text6869-2-0)
  {$\sqrt{\frac{2}{5}}\ket{000}$};

\path[fill=black,line join=miter,line cap=butt,line width=0.800pt]
  (144.0000,165.3622) node[above right] (text6869-2-0-3)
  {$\frac{1}{\sqrt{5}}\ket{010}$};

\path[fill=black,line join=miter,line cap=butt,line width=0.800pt]
  (196.5000,165.8622) node[above right] (text6869-2-0-3-0)
  {$\frac{1}{\sqrt{5}}\ket{100}$};

\path[fill=black,line join=miter,line cap=butt,line width=0.800pt]
  (296.5000,165.8622) node[above right] (text6869-2-0-3-0-2)
  {$\frac{1}{\sqrt{5}}\ket{110}$};

\path[fill=black,line join=miter,line cap=butt,line width=0.800pt]
  (6.0000,253.0000) node[above right] (text6869-2-0-3-0-1)
  {$\frac{1}{\sqrt{5}}\ket{000}$};

\path[fill=black,line join=miter,line cap=butt,line width=0.800pt]
  (61.0000,253.0000) node[above right] (text6869-2-0-3-0-7)
  {$\frac{1}{\sqrt{5}}\ket{001}$};

\path[fill=black,line join=miter,line cap=butt,line width=0.800pt]
  (135.5000,251.3622) node[above right] (text6869-2-0-3-0-20)
  {$\frac{1}{\sqrt{5}}\ket{010}$};

\path[fill=black,line join=miter,line cap=butt,line width=0.800pt]
  (207.5000,251.3622) node[above right] (text6869-2-0-3-0-3)
  {$\frac{1}{\sqrt{5}}\ket{101}$};

\path[fill=black,line join=miter,line cap=butt,line width=0.800pt]
  (283.0000,251.3622) node[above right] (text6869-2-0-3-0-5)
  {$\frac{1}{\sqrt{5}}\ket{110}$};

\path[fill=black,line join=miter,line cap=butt,line width=0.800pt]
  (159.9889,64.9541) node[above right] (text6869-2-2) {$\ket{000}$};

\end{tikzpicture}
}\label{fig:bsqdd5}}
 \subfloat[Final BSQDD]{{\scalefont{0.85}

\begin{tikzpicture}[y=0.80pt, x=0.80pt, yscale=-1.000000, xscale=1.000000, inner 
sep=0pt, outer sep=0pt]
\path[draw=black,line join=miter,line cap=butt,miter limit=4.00,line
  width=0.560pt] (309.7885,397.0240) circle (0.4233cm);

\path[draw=black,line join=miter,line cap=butt,miter limit=4.00,line
  width=0.600pt] (309.7886,372.0241) -- (309.7886,382.0241);

\path[draw=black,line join=miter,line cap=butt,miter limit=4.00,line
  width=0.600pt] (294.0854,397.0241) -- (254.0854,427.0241);

\path[draw=black,line join=miter,line cap=butt,miter limit=4.00,line
  width=0.600pt] (324.7886,397.0241) -- (384.7886,427.0241);

\path[draw=black,line join=miter,line cap=butt,miter limit=4.00,line
  width=0.648pt] (253.2338,438.8065) ellipse (0.7769cm and 0.3092cm);

\path[draw=black,line join=miter,line cap=butt,miter limit=4.00,line
  width=0.639pt] (387.1948,438.3151) ellipse (0.7338cm and 0.3180cm);

\path[draw=black,line join=miter,line cap=butt,miter limit=4.00,line
  width=0.600pt] (225.7886,437.0241) .. controls (217.8087,447.3700) and
  (230.7886,467.0241) .. (240.7886,482.0241);

\path[draw=black,line join=miter,line cap=butt,miter limit=4.00,line
  width=0.600pt] (279.7886,437.0241) .. controls (286.1055,452.2807) and
  (274.7886,472.0241) .. (264.7886,482.0241);

\path[draw=black,line join=miter,line cap=butt,miter limit=4.00,line
  width=0.673pt] (252.8578,490.6912) ellipse (1.0797cm and 0.2394cm);

\path[draw=black,line join=miter,line cap=butt,miter limit=4.00,line
  width=0.668pt] (387.1746,489.7604) ellipse (0.9931cm and 0.2568cm);

\path[draw=black,line join=miter,line cap=butt,miter limit=4.00,line
  width=0.600pt] (223.6389,496.8567) -- (201.2841,522.6379);

\path[draw=black,line join=miter,line cap=butt,miter limit=4.00,line
  width=0.600pt] (274.6767,497.9951) -- (298.9961,523.2517);

\path[draw=black,line join=miter,line cap=butt,miter limit=4.00,line
  width=0.600pt] (250.7886,498.0241) -- (250.7886,523.0241);

\path[draw=black,line join=miter,line cap=butt,miter limit=4.00,line
  width=0.600pt] (402.9470,498.1624) -- (417.9470,523.1624);

\path[draw=black,line join=miter,line cap=butt,miter limit=4.00,line
  width=0.600pt] (371.1055,497.5486) -- (356.1055,522.5486);

\path[draw=black,line join=miter,line cap=butt,miter limit=4.00,line
  width=0.600pt] (413.7886,437.0241) .. controls (420.1055,452.2807) and
  (413.7886,472.0241) .. (403.7886,482.0241);

\path[draw=black,line join=miter,line cap=butt,miter limit=4.00,line
  width=0.600pt] (360.7886,437.0241) .. controls (352.8087,447.3700) and
  (365.7886,467.0241) .. (375.7886,482.0241);

\path[fill=black,line join=miter,line cap=butt,line width=0.800pt]
  (298.5000,369.8622) node[above right] (text4787) {$\ket{000}$};

\path[fill=black,line join=miter,line cap=butt,line width=0.800pt]
  (356.5000,410.8622) node[above right] (text4791)
  {$\sqrt{\frac{2}{5}}\ket{100}$};

\path[fill=black,line join=miter,line cap=butt,line width=0.800pt]
  (227.5000,410.3622) node[above right] (text4791-0)
  {$\sqrt{\frac{3}{5}}\ket{000}$};

\path[fill=black,line join=miter,line cap=butt,line width=0.800pt]
  (309.7885,397.0240) node[] (text4813) {$\mathtt{R(\theta)}$};

\path[fill=black,line join=miter,line cap=butt,line width=0.800pt]
  (253.2338,438.8065) node[] (text4817)
  {$\mathtt{CR(\alpha)^0_{32}}$};

\path[fill=black,line join=miter,line cap=butt,line width=0.800pt]
  (387.1948,438.3151) node[] (text4817-3) {$\mathtt{CW_{32}}$};

\path[fill=black,line join=miter,line cap=butt,line width=0.800pt]
  (252.8578,490.6912) node[] (text4839) {$\mathtt{CCW^{00}_{321}}$};

\path[fill=black,line join=miter,line cap=butt,line width=0.800pt]
  (387.1746,489.7604) node[] (text4839-0)
  {$\mathtt{CCX^{10}_{321}}$};

\path[fill=black,line join=miter,line cap=butt,line width=0.800pt]
  (179.5000,470.3622) node[above right] (text4791-0-2)
  {$\sqrt{\frac{2}{5}}\ket{000}$};

\path[fill=black,line join=miter,line cap=butt,line width=0.800pt]
  (278.0000,471.8622) node[above right] (text4791-0-2-1)
  {$\frac{1}{\sqrt{5}}\ket{010}$};

\path[fill=black,line join=miter,line cap=butt,line width=0.800pt]
  (336.0000,469.8622) node[above right] (text4791-0-2-1-2)
  {$\frac{1}{\sqrt{5}}\ket{100}$};

\path[fill=black,line join=miter,line cap=butt,line width=0.800pt]
  (416.5000,470.8622) node[above right] (text4791-0-2-1-2-2)
  {$\frac{1}{\sqrt{5}}\ket{110}$};

\path[fill=black,line join=miter,line cap=butt,line width=0.800pt]
  (407.0000,541.3622) node[above right] (text4791-0-2-1-2-2-7)
  {$\frac{1}{\sqrt{5}}\ket{110}$};

\path[fill=black,line join=miter,line cap=butt,line width=0.800pt]
  (346.0000,541.3622) node[above right] (text4791-0-2-1-2-2-9)
  {$\frac{1}{\sqrt{5}}\ket{101}$};

\path[fill=black,line join=miter,line cap=butt,line width=0.800pt]
  (289.5000,541.3622) node[above right] (text4791-0-2-1-2-2-9-2)
  {$\frac{1}{\sqrt{5}}\ket{010}$};

\path[fill=black,line join=miter,line cap=butt,line width=0.800pt]
  (240.0000,541.3622) node[above right] (text4791-0-2-1-2-2-9-2-9)
  {$\frac{1}{\sqrt{5}}\ket{001}$};

\path[fill=black,line join=miter,line cap=butt,line width=0.800pt]
  (177.0000,541.3622) node[above right] (text4791-0-2-1-2-2-9-2-9-3)
  {$\frac{1}{\sqrt{5}}\ket{000}$};

\end{tikzpicture}
}\label{fig:bsqdd6}}
 \vspace*{2em} \\\subfloat[Corresponding quantum array]{
 $\Qcircuit @C=1.em @R=.7em{ 
&&\ustick{\ket{z}}
\gategroup{2}{4}{4}{4}{1.em}{--}\gategroup{1}{3}{5}{3}{.0em}{--}&&
\ustick{\ket{\psi_3}}\gategroup{1}{5}{5}{5}{.0em 
}{--}&&&&\ustick{\ket{\psi_2}}\gategroup{1}{9}{5}{9}{.0em 
}{--}\gategroup{2}{7}{4}{8}{2.4em}{--}&&&&\ustick{\ket{\psi_1}}
\gategroup{1}{13}{5}{13}{.0em}{--}\gategroup{2}{11}{4}{12}{1.em}{--}\\ 
&\lstick{\ket{0}}&\qw&\gate{\mathtt{R}(\theta)}&\qw&\qw&\ctrlo{1}&\ctrl{1}
&\qw&\qw&\ctrlo{1}&\ctrl{1}&\qw\\ 
&\lstick{\ket{0}}&\qw&\qw&\qw&\qw&\gate{\mathtt{R}(\alpha)}&\gate{\mathtt{W}} 
&\qw&\qw&\ctrlo{1}&\ctrlo{1}&\qw\\ 
&\lstick{\ket{0}}&\qw&\qw&\qw&\qw&\qw&\qw&\qw&\qw&\gate{\mathtt{W}}
&\gate{\mathtt{X}}&\qw\\
 &&& G_3&&&G_2&&&&G_1&&
 } $
 }
 \caption{BSQDD to obtain the state 
$\ket{\psi_1}=\sqrt{\frac{1}{5}}(\ket{000}+\ket{010}+\ket{110}+\ket{001}+ 
\ket{101})$. The two BSQDDs of the Fig. are equivalent.}
 \label{fig:learn1}
 \end{figure}
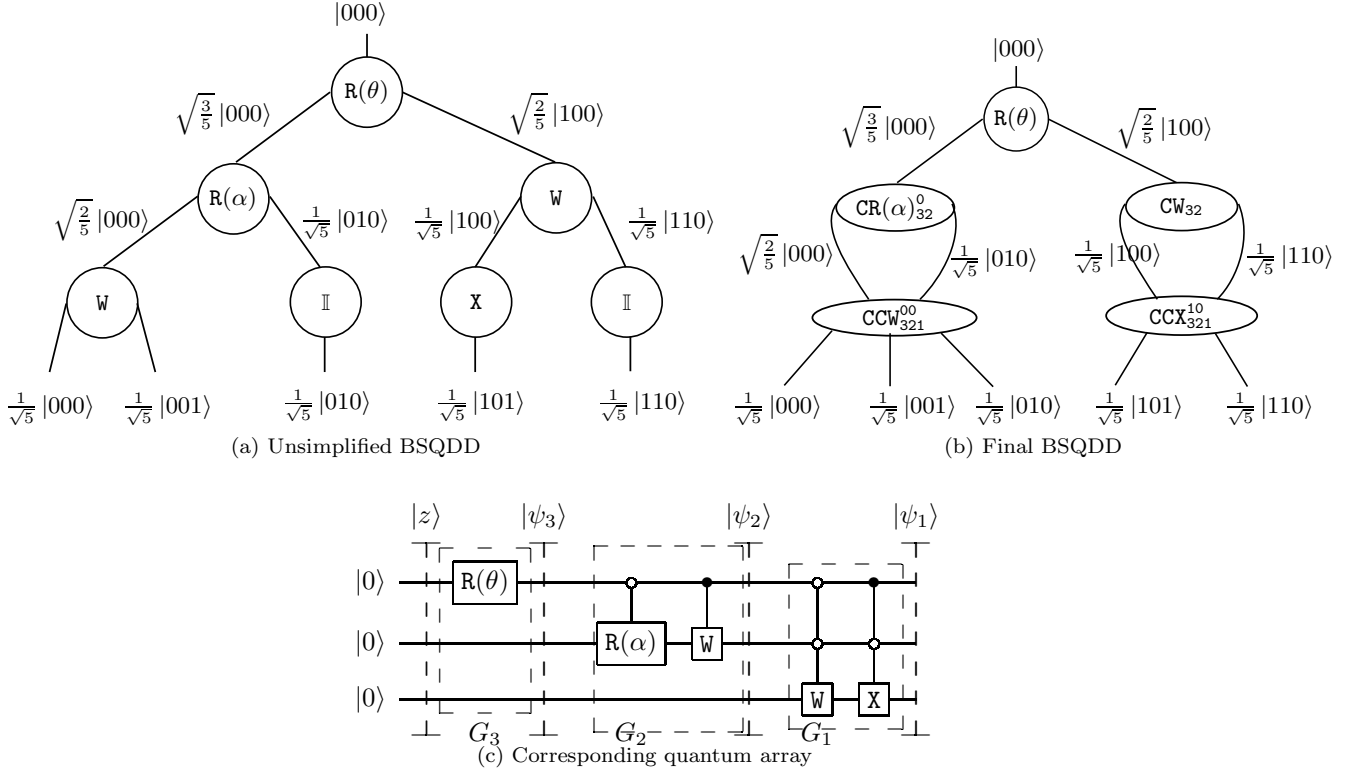
 
\end{example}

\subsection{Quantum retrieving algorithms}\label{sub_Sec:retri}
We use two quantum retrieving algorithms that work together but with different 
learning approaches.

\subsubsection{Linear QAM algorithm}
The linear QAM retrieving algorithm given by Algorithm \ref{alg:linear} is a
slightly modified version of \textbf{QAM-C1} that we have proposed in ref.
\cite{tchapet2012}. It uses an \emph{exclusion learning approach}. That 
approach states that the
system must be in the superposition of all the possible states, except the 
pattern states. Let us consider $M$ as a set of patterns and $m$ the number 
of patterns of length $n$, then
\begin{equation}
\ket{\Psi}=\frac{1}{\sqrt{N-m}}\sum^{N-1}_{x\notin M}\ket{x},\,N=2^n.
\label{eq:Psi}
\end{equation}

\begin{algorithm}[htpb]
\caption{Linear QAM retrieving algorithm with distributed query for diagnosis}
\label{alg:linear}
\begin{algorithmic}[1]
\STATE \label{etap2} Apply the oracle operator
$\mathcal{O}$ to the register;
\STATE Apply the diffusion operator $\mathcal{D}$ to the
register;
\IF{$\Lambda>1$}
\STATE \label{etap3} Apply the operator $\mathcal{I}_M$
to the register;
\STATE \label{etap4} Apply the diffusion 
operator $\mathcal{D}$ to the register;
\FOR{$1\leq i \leq \Lambda-1$}
\STATE \label{etap5} Apply the oracle operator
$\mathcal{O}$ to the register;
\STATE \label{etap6} Apply the diffusion operator
$\mathcal{D}$ to the register;
\STATE $i=i+1$;
\ENDFOR
\ENDIF
\STATE Observe the system.
\end{algorithmic}
\end{algorithm}

In the Algorithm \ref{alg:linear},
\begin{itemize}
 \item $\mathcal{O}$ is the oracle operator that inverts the phase of the
query state $\ket{Req^p}$,
 \begin{align}
 \mathcal{O}& =\mathbb{I}-(1-e^{i\pi})\ket{Req^p}\bra{Req^p},\\
  \mathcal{O} &:a_x\mapsto a_x-2Req^p_x\left(\sum^{2^{n}-1}_{x=0}
(Req^p_x)^{*}a_x\right),
\label{eq:O}
\end{align}
where $a_x$ is the probability amplitude of the state $\ket{x}$. 
The distributed query $\ket{Req^p}$ is in the following superposed states
\begin{equation}
\ket{Req^p}=\sum^{N-1}_{x=0}Req^p_x\ket{x},
\end{equation}
where $Req^p_x$ obey to binomial distribution
\begin{equation}
\|Req^p_x\|^{2}=a^{d_H(p,x)}(1-a)^{n-d_H(p,x)}.
\label{eq:re}
\end{equation}
In equation (\ref{eq:re}),
\begin{itemize}
\item $p$ marks the state $\ket{p}$ that is referred as the query centre;
\item $a\in]0,\frac{1}{2}[$ is an arbitrary value that tunes the width of the
distribution;
\item the \textbf{Hamming distance} $d_H(p,x)=|p-x|$ between binary strings $p$ 
and $x$ is an important tool that gives the correlation between input and
output;
\item the amplitudes are such that $\sum_x\|Req^p_x\|^{2}=1$.
\end{itemize}

\item $\mathcal{D}$ is the diffusion operator that inverts the probability
amplitude of the states of $\ket{\Psi}$ over their average amplitude and for the
others over the value $0$:
 \begin{align}
  \mathcal{D}& =(1-e^{i\pi})\ket{\Psi}\bra{\Psi}-\mathbb{I},\\
 \mathcal{D} &:a_x\mapsto 2m_x\left(\sum^{N-1}_{x=0}m^{*}_xa_x\right)-a_x,
\label{eq:D}
\end{align}
where $m_x$ is the probability amplitude of a state of $\ket{\Psi}$.

\item $\Lambda$ is the number of iterations that yields the maximal value of
amplitudes, which must be as far as possible nearest to an integer,
\begin{equation}
 \Lambda=T(\frac{1}{4}+\alpha),\,T=\frac{2\pi}{\omega},\,\alpha\in\mathbb{N},
\label{eq:lambda}
\end{equation}
with the frequency of Grover
\begin{equation}
\label{equaB}
 \omega=2\arcsin{B},\,B=\frac{1}{\sqrt{N-m}}\sum^{N-1}_{x=0,x\notin M}Req^p_x.
\end{equation}

\item According to the approach \textbf{QAM-C1} of the ref.
\cite{tchapet2012} $\mathcal{I}_{M}$ inverts only the phase of the memory
pattern states as in the model of Ventura,
\begin{subequations}
\begin{equation}
 \mathcal{I}_{M} =\mathbb{I}-(1-e^{i\pi})\ket{\varphi}\bra{\varphi},\,
\ket{\varphi}\bra{\varphi}=\sum_{x\in M}\ket{x}\bra{x},
\end{equation}
\begin{equation}
 \mathcal{I}_{M}:a_{x}\mapsto
\begin{cases}
-a_{x}\text{ if }\ket{x}\in M\\
a_{x}\text{ if not.}
\end{cases}
\end{equation}
\end{subequations}
\end{itemize}

The linear QAM algorithm is used here as main algorithm given that it increases 
the probability amplitude of the searched disease.

\subsubsection{Non-linear QAM algorithm}
The non-linear QAM retrieving algorithm given by Algorithm \ref{alg:non-linear}
is formally the same with the one we have proposed in ref.
\cite{tchapet2016} (see that ref. for more details) that is an improved
version of the one given by \cite{Abrams1998}. It uses \emph{inclusion learning
approach}.
\begin{algorithm}[H]
\caption{Non-linear QAM retrieving algorithm for diagnosis}
\label{alg:non-linear}
\begin{algorithmic}[1]
\STATE Apply the oracle operator $\mathtt{U}_f$, with $f$ a function
\STATE \textbf{repeat} $(c-r)$ times  step (\ref{alg:begin}) to step
(\ref{alg:end}) (i.e., one time per qubit of the first register starting from
$(r+1)^{th}$ qubit with the flag qubit)
\STATE\hspace{1em}\label{alg:begin} Apply the unitary operator 
$\mathtt{U}$
\STATE \begin{enumerate}
	\item Apply the non-linear operator $\mathtt{NL}^-$
	\item Apply the non-linear operator $\mathtt{NL}^+$
\end{enumerate}
\STATE\hspace{1em}\label{alg:end} Apply the Hadamard operator 
$\mathtt{W}$ on
the qubit of the first register and the $\mathtt{NOT}$ operator 
$\mathtt{X}$ on the flag
qubit
\STATE Observe the flag qubit
\end{algorithmic}
\end{algorithm}

Algorithm \ref{alg:non-linear} is used on the subspace of signs and symptoms 
and 
allows to
know if signs and symptoms related to a particular disease exist in the 
database. In the
Algorithm \ref{alg:non-linear}, 
\begin{itemize}
 \item $n$ is the number of qubits of the first register.
 \item $p\leq2^n$ is the number of stored patterns.
 \item $q\leq p$ is the number of stored patterns if the values of $t$ qubits
 are known (i.e. $t$ qubits have been measured or are already disentangled to
 others or the oracle acts on a subspace of $(n-t)$ qubits).
\item $c=\mathtt{ceil}(\log_2{q})$, that is the least integer greater or equal 
to $\log_2{q}$.
\item $m\leq q$ is the number of values $x$ for which $f(x)=1$.
\item $r=\mathtt{int}(\log_2{m})$ is the integer part of $\log_2{m}$.
\end{itemize}

\section{Database: signs and symptoms of the four tropical 
diseases}\label{sec:dtb}

\subsection{Short description of the four diseases}\label{subsec:dis}

Based on \cite{epilly,manson,ecn,msf} we are going to give a briefly 
description of the four tropical diseases of our study. Appendix 
\ref{sec:signs} 
gives their signs and symptoms.

\begin{description}
\item[Malaria] is a life-threatening disease caused by protozoan parasites of
the genus \emph{Plasmodium}. The parasite is generally transmitted from one
human to another through the bite of infected females Anopheles mosquitoes. Five
species infect humans by entering the bloodstream: \emph{P. knowlesi, P. ovale, 
P. malariae, P. vivax and P. falciparum}. The last two ones, \emph{P. 
vivax}
and\emph{ P. falciparum}, are the greatest threat because they affect a 
greater proportion of the red blood cells than the others. \emph{P. falciparum} 
is the most prevalent malaria parasite on the African continent. It is 
generally responsible for
most malaria-related deaths. \emph{P. vivax} has a wider distribution
than \emph{P. falciparum}, and predominates in many countries outside Africa.
About 3.2 billion people, almost half of the world population, are exposed to 
malaria. Sub-saharan Africa carries a disproportionately high share of the 
global malaria burden. In 2015, the region was home of $89\% $ of malaria cases 
and $91\%$ of malaria deaths.

\item[Typhoid fever], also known as salmonellosis,  is a life-threatening
illness caused by the bacterium \emph{Sallmonella typhi}, also known as
\emph{Salmonella enterica} serotype typhi, growing in the intestines and blood.
Dirt (poor sanitation and poor hygiene) is the main cause of transmission of the
disease. Contaminated food and unsafe water are the main vectors. Typhoid 
fever
remains a serious worldwide threat, especially in developing countries. The 
estimated cases of this disease are nearby 16-33 millions each year; this leads
to more than half million death cases. The disease is endemic in India, 
Southeast Asia,
Africa, South America and many other areas.

\item[Yellow fever] is an acute viral hemorrhagic disease caused by a
\emph{Flavivirus}. The virus is transmitted through the bite of an infected
female mosquito (\emph{Aedes aegypti}). The virus causes deterioration of the
liver. There are an estimated 200,000 cases of yellow fever, causing 30,000
worldwide deaths each year, with $ 90\% $ occurring in Africa. Forty-four
endemic countries in Africa and Latin America, with a combined population of
over 900 millions, are at risk. In Africa, an estimated 508 millions people live
in 31 countries exposed to the disease. The remaining population at risk are 
in 13 countries in
Latin America. Bolivia, Brazil, Colombia, Ecuador and Peru are at greatest
risk.

\item[Dengue fever] is a painful, debilitating mosquito-borne tropical disease
caused by \emph{dengue virus}. The virus is transmitted by several species of
mosquito within the genus \emph{Aedes}, principally \emph{Aedes aegypti}. WHO
has classified dengue as one of the neglected tropical diseases and reported the
resurgence of the disease \cite{neglige}. Dengue is common in more than 110
countries. It infects 50 to 528 millions people worldwide a year, leading to 
half million hospitalisations, and approximately 25,000 deaths. Cases of the
disease have been reported at least in 22 countries in Africa; but it is likely 
present in all of them with $
20\% $ of the population at risk. This makes it one of the most common
vector-borne diseases worldwide.
\end{description}

\subsection{Description of the database}\label{subsec:data}

Our database contains signs and symptoms of four tropical diseases: malaria, 
typhoid
fever, dengue fever and yellow fever (see Appendix \ref{sec:signs}). There are
seventeen signs and symptoms specific to malaria, eighteen to typhoid fever, 
twenty-three to 
dengue
and seventeen to
yellow fever. Some signs and symptoms are common to different
diseases, not only the four, such as fever and headache for example, thereby 
our 
database contains ninety-five signs and symptoms. We need six qubits for the 
computation because 
each symptom is labelled with a number in its binary form. Therefore we can 
compute $2^6$ signs and symptoms per disease. In the subspace of disease we 
label the 
signs and symptoms in decimal form by starting with $0$.
Furthermore, we need four qubits to label the ten groups of diseases presented 
by the Tab. \ref{tab:indexDisease} that are used in the linear QAM algorithm. 
There 
is one group per individual disease (N\textdegree{} 1-4); two groups 
corresponding to common signs and symptoms to malaria and typhoid fever or 
yellow fever
(N\textdegree{} 5-6); two groups corresponding to common signs and symptoms to 
yellow 
fever and typhoid fever or dengue (N\textdegree{} 7- 8); one group 
corresponding to common signs and symptoms to malaria, yellow fever and dengue 
(N\textdegree{} 9); and finally one group
corresponding to common signs and symptoms to each of the four diseases. The 
last 
can be
found in other diseases such as headache, fever and abdominal pain 
(N\textdegree 10).
For the non-linear QAM algorithm we need four flag qubits, one for each 
disease. 
It appears that we need a register that contains $n=14$ qubits, six for 
signs and symptoms, four for diseases for each QAM algorithm. So, there are two 
output 
registers. The labels of the output qubits for the linear QAM algorithm are 
mentioned above. The other possibilities are pointed to be other diseases and 
signs and symptoms in our model. Two sets of four qubits for output are needed 
due to the 
fact that the learning process of the linear QAM algorithm is different from 
the one used for the non-linear QAM algorithm. For the non-linear QAM 
algorithm each flag qubit is associated to one disease. That qubit allows to 
know if a symptom specific to a particular disease is present. Therefore value 
$0$ means that this particular disease is not present, whereas value $1$ means 
that it is present. Nevertheless, if the values of that four qubits are $0$, it 
means that none specific symptom is introduced.

\begin{table}[hbtp]
\centering
\begin{tabular}{llc}\hline
\textbf{N\textdegree} &\textbf{Group of diseases by signs and symptoms} & 
\textbf{Label}\\\hline
1 & Malaria & $\ket{0001}$\\
2 & Typhoid fever & $\ket{0010}$\\
3 & Yellow fever & $\ket{0100}$\\
4 & Dengue & $\ket{1000}$\\
5 & Malaria + Typhoid fever & $\ket{0011}$\\
6 & Malaria + Yellow fever & $\ket{0101}$\\
7 & Yellow fever + Typhoid fever & $\ket{0110}$\\
8 & Yellow fever + Dengue & $\ket{1100}$\\
9 & Malaria + Yellow fever + Dengue & $\ket{1101}$\\
10 & Other diseases & $\ket{1111}$\\\hline
\end{tabular}
\caption{Groups of diseases by signs and symptoms and their labels in binary 
form for 
the linear algorithm. The hamming distance between the label of two groups of
diseases is equal to $1$ when the signs and symptoms are common to these two 
groups of
diseases and is equal to $2$ otherwise. The group N\textdegree{} 10 or
\emph{Other diseases} is devoted to signs and symptoms that are common to each
$4$ diseases and that can also occur in other groups of diseases not 
mentioned here. We point out the fact that other labels are also
consider to be \emph{Other diseases.}}
\label{tab:indexDisease}
\end{table}

For the linear QAM retrieving algorithm, the determination of the number of
iterations is mentioned above.

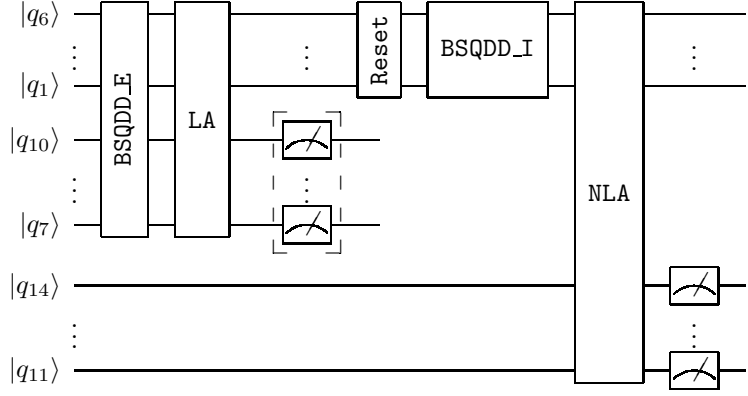
\begin{figure}[H]
\[\Qcircuit @C=1.0em @R=.9em{ 
	&\lstick{\ket{q_6}}&\multigate{5}{\rotatebox{90}{$\mathtt{BSQDD\_E}$}}
	&\multigate{5}{\mathtt{LA}}
	&\qw&\qw&\multigate{2}{\rotatebox{90}{$\mathtt{Reset}$}}&\multigate{2}{\mathtt{
			BSQDD\_I}}
	&\multigate{8}{\mathtt{NLA}} &\qw&\qw\\
	&\vdots\gategroup{4}{6}{6}{6}{.7em}{--}&&&&\vdots&&&&\vdots&\\ 
	&\lstick{\ket{q_1}}&\ghost{\rotatebox{90}{$\mathtt{BSQDD\_E}$}}&\ghost{\mathtt{
			LA}}
	&\qw&\qw&\ghost{
		\rotatebox{90}{$\mathtt{Reset}$}}&\ghost{\mathtt{BSQDD\_I}}
	&\ghost{\mathtt{NLA}} &\qw&\qw\\ 
	&\lstick{\ket{q_{10}}}&\ghost{\rotatebox{90}{$\mathtt{BSQDD\_E}$}}&\ghost{
		\mathtt{LA}}
	&\qw&\meter&\qw&&&&\\
	&\vdots&&&&\vdots&&&&&\\ 
	&\lstick{\ket{q_{7}}}&\ghost{\rotatebox{90}{$\mathtt{BSQDD\_E}$}}&\ghost{\mathtt
		{LA}}
	&\qw&\meter&\qw&&&&\\ 
	&\lstick{\ket{q_{14}}}&\qw&\qw&\qw&\qw&\qw&\qw&\ghost{\mathtt{NLA}}&\meter&\qw\\
	&\vdots&&&&&&&&\vdots&\\
	&\lstick{\ket{q_{11}}}&\qw&\qw&\qw&\qw&\qw&\qw&\ghost{\mathtt{NLA}}&\meter&\qw\\
}\]
 \caption{Schematic structure of the QAMDiagnos. $\mathtt{BSQDD\_E}$ is 
the learning part of the linear QAM algorithm, whereas, after reinitialisation,
 $\mathtt{BSQDD\_I}$ is the learning part of the non-linear QAM algorithm. 
$\mathtt{LA}$ is the set of gates simulating the linear QAM algorithm. 
$\mathtt{NLA}$ is a repeated non-linear QAM algorithm, once for each disease. 
Each flag qubit from $\ket{q_{11}}$ to $\ket{q_{14}}$ is devoted to one 
disease.}
 \label{fig:QAM}
\end{figure}

Labelling all the fourteen qubits from $\ket{q_1}$ to $\ket{q_{14}}$, the 
entire QAM looks as illustrated in Fig. \ref{fig:QAM}.

\section{Simulations results and discussion}\label{sec:Sim}
To design the linear part of the QAMDiagnos, we take into account the four following
facts:
\begin{enumerate}
\item The arbitrary value that regulates the width distribution around the 
chosen centre is $a = 0.4999$ ;
\item The centre of the query is $\ket{0000}$ because we assume that there is 
no indication about disease;
\item The difference between the number of iterations $\Lambda$ and the 
nearest integer must be less than $0.1$;
\item Not more than six signs and symptoms can be chosen.
\end{enumerate}

We give in the next Tabs. \ref{tab:singleinfect} and \ref{tab:multiinfect} the
average probabilities of correct recognition ($P_c$). We take the average due to
the convention we chose to label signs and symptoms, and diseases. Therefore,
\emph{sensibility} is the conditional probability to have a correct recognition
if the disease is present. So, as we can obtain from our QAM probabilities of
good recognition, we can see that they look like sensibilities.

\begin{itemize}
 \item \textbf{For single infection}
 
 We obtain sensibilities according to the definition.

\begin{table}[H]
\centering
\begin{footnotesize}{
\newcommand{\mc}[3]{\multicolumn{#1}{#2}{#3}}
\definecolor{tcA}{rgb}{0.627451,0.627451,0.643137}
\begin{center}
\begin{tabular}{p{2cm}ccccccc||cccc}\\\hline
	& \mc{6}{c}{\textbf{Number of signs and symptoms}}  & \\\cline{2-7}
	\textbf{Disease}& \textbf{1} & \textbf{2} 
	& \textbf{3} & \textbf{4} & \textbf{5} & \textbf{6}& &$q_{11}$& 
$q_{12}$& $q_{13}$& $q_{14}$\\\hline
	Malaria & $93.8911$  & $94.5059$  & $97.5402$ & 
	$96.4843$  & $99.9043$  & $96.5960$ &Without &$1$&$0$&$0$&$0$\\
	Typhoid fever & $93.8911$  & $94.5059$  & $97.5402$ & 
	$96.4843$  & $99.9043$  & $96.5960$
	& other&$0$&$1$&$0$&$0$\\
	Yellow fever & $93.8911$  & $94.5059$  & $97.5402$ & 
	$96.4843$  & $99.9043$  & $96.5960$
	&&$0$&$0$&$1$&$0$\\
	Dengue & $93.8904$  & $94.5052$  & $97.5397$ 
	& $96.4837$  & $99.9041$  & $96.5954$ & 
signs and symptoms&$0$&$0$&$0$&$1$\\\hline
	Malaria & $47.4418$  & $65.0912$  & $72.4296$  & $79.9350$  & 
	$80.5407$ & 
	- & With 1&$1$&$0$&$0$&$0$\\
	Typhoid fever& $47.4418$  & $65.0912$  & $72.4296$  & $79.9350$  & 
	$80.5407$ & 
	-&other&$0$&$1$&$0$&$0$\\
	Yellow fever & $47.4418$  & $65.0912$  & $72.4296$  & $79.9350$  & 
	$80.5407$ & 
	- &&$0$&$0$&$1$&$0$\\
	Dengue & $47.4411$  & $65.0905$  & $72.4289$ 
	&  $79.9346$  & $80.5400$  & - &sign or symptom&$0$&$0$&$0$&$1$\\\hline
	Malaria & $32.6307$ & $48.3689$ & 
$59.9605$ & 	$64.4826$  & - & 
	- & With 2&$1$&$0$&$0$&$0$\\
	Typhoid fever &$32.6307$ & $48.3689$ & 
$59.9605$ & 	$64.4826$  & - & 
	- &other&$0$&$1$&$0$&$0$\\
	Yellow fever & $32.6307$ & $48.3689$ & 
$59.9605$ & 	$64.4826$  & - & 
	- &&$0$&$0$&$1$&$0$\\
	Dengue & $32.6302$ & $48.3683$ & $59.9600$
	&  $64.4819$  & -  & - &signs and symptoms&$0$&$0$&$0$&$1$\\\hline
	Malaria & $24.3022$ & $39.9808$ & $48.4219$  & -  & - & 
	- & With 3&$1$&$0$&$0$&$0$\\
	Typhoid fever & $24.3022$ & $39.9808$ & $48.4219$  & -  & - & 
	-  &other&$0$&$1$&$0$&$0$\\
	Yellow fever & $24.3022$ & $39.9808$ & $48.4219$  & -  & - & 
	-  &&$0$&$0$&$1$&$0$\\
	Dengue & $24.3018$ & $39.9803$ & $48.4213$ 
	&  -  & -  & - &signs and symptoms&$0$&$0$&$0$&$1$\\\hline
\end{tabular}
\end{center}
}%
\end{footnotesize}
\caption{Average probabilities $P_c$ according to the number of signs and
	symptoms related to a disease in case of a single infection for the linear part
	of the QAM. For the non-linear part, each corresponding flag qubit has its value
	equal to $1$. Here ``other signs and symptoms'' are fever, headache and
	abdominal pain.} \label{tab:singleinfect}
\end{table}

 \item \textbf{For polyinfection}
 
 We also obtain sensibilities according to the definition.
 
 \begin{table}[H]
\centering
\begin{footnotesize}{
\newcommand{\mc}[3]{\multicolumn{#1}{#2}{#3}}
\definecolor{tcA}{rgb}{0.627451,0.627451,0.643137}
\begin{center}
\begin{tabular}{p{3cm}ccccp{2cm}}\\\hline
	& \mc{4}{c}{\textbf{Number of symptom of malaria}}  & 
\\\cline{2-5}
	\textbf{Disease}&\textbf{0}& \textbf{1} & \textbf{2} 
	& \textbf{3}& \\\hline
	Malaria & $0.1577$  & $24.2913$ & $39.9633$  & $48.4079$ &With 1 sign\\
	Typhoid fever  & $32.6167$ & $24.2913$ & $19.9844$  
	& $16.2842$  &  or symptom\\
	Malaria + Typhoid f.  & $65.0886$ & $48.3653$ 
	& $39.9685$  & $32.3538$  & of Typhoid f.\\\hline
	Malaria& $0.2283$ & $19.9844$ & $32.3460$  & - &With 2 signs\\
	Typhoid fever& $48.3543$ & $39.9633$ & $32.3460$  
	& -  &  and symptoms\\
	Malaria + Typhoid f.  & $48.3653$ & $39.9685$ 
	& $32.3538$  & -  &  of Typhoid f.\\\hline
	Malaria& $0.0055$ & $16.2842$ & -  & - &With 3 signs\\
	Typhoid fever& $59.9422$ & $48.4079$ & -  
	& -  & and symptoms\\
	Malaria + Typhoid f.  & $39.9685$ & $32.3538$ 
	& -  & -  &  of Typhoid f.\\\hline
	Malaria& $0.2224$ & - & -  & - &With 4 signs\\
	Typhoid fever& $64.4697$ & - & -  
	& -  &  and symptoms\\
	Malaria + Typhoid f.  & $32.3538$ & - 
	& -  & -  & of Typhoid f.\\\hline\hline
	$q_{11}$&$0$&$1$&$1$&$1$&\\
	$q_{12}$&$1$&$1$&$1$&$1$&\\
	$q_{13}$&$0$&$0$&$0$&$0$&\\
	$q_{14}$&$0$&$0$&$0$&$0$&\\\hline
\end{tabular}
\end{center}
}%
\end{footnotesize}
\caption{Probabilities $P_c$ according to the number of signs and symptoms
	related to diseases in case of a polyinfection for the linear part of the QAM.
	For the non-linear part, each corresponding flag qubit has its value equal to
	$1$. Here ``common sign or symptom`` is the one that is common to the two diseases. Here
	we have chosen two common signs and symptoms.}
\label{tab:multiinfect}
\end{table}

As we can see on the Tabs. \ref{tab:singleinfect} and \ref{tab:multiinfect} the
QAM can collapse to a state representing single infection or polyinfection. That
is the QAM can distinguish single from polyinfection. That distinction is
possible with the lowest (one) or the highest (six) number of particular signs
and symptoms of a disease, but it is better in the last case. When the ``other
signs and symptoms'' are inserted, the QAM can also do the distinction. As we
observe on the Tab. \ref{tab:otherinfect} the ``other signs and symptoms'' are
not related to a particular disease. Therefore, the non-linear part of the QAM
completes or corrects the results of the linear part.

\begin{table}[H]
\centering
\begin{footnotesize}{%
\newcommand{\mc}[3]{\multicolumn{#1}{#2}{#3}}
\begin{center}
\begin{tabular}{lccc}\hline
	&\mc{3}{c}{\textbf{Number of signs and symptoms}}\\\cline{2-4}
	\textbf{Disease} & \textbf{1} & \textbf{2} & \textbf{3}\\\hline
	Other diseases & $96.38$  & $96.73$ &  $98.53$ \\\hline
\end{tabular}
\end{center}
}%
\end{footnotesize}
\caption{Probabilities $P_c$ according to the number of ``other signs and
	symptoms'' for linear part of the QAM. For the non-linear part, each flag qubit
	has its value equal to $0$. The QAM does not associate these signs and symptoms
	to one of the four diseases.} \label{tab:otherinfect}
\end{table}
\end{itemize}

We also evaluate the specificity of the QAM. \emph{Specificity} is the
conditional probability to have a correct recognition if the disease is not
present. In other words it is the ability of the memory to distinguish healthy
people from non-healthy ones. For our QAM, to acquire it we use the $448$ signs
and symptoms without any relation with the four diseases and take the average of
probabilities to obtain ``Other diseases'' as a result. We obtain the Tab.
\ref{tab:otherinfect} and assume that the specificity is $\mathbf{96.38\%}$.

\section{QAMDiagnos: desktop and smart-phone 
GUI}\label{sec:over}

All the simulations and results were made by encoding the algorithms in
\verb|C++| language. A multi-platform friendly graphical user interface (GUI) of
our QAMDiagnos is designed for the medical staff. (see figures of supplementay
material for more details). It is developed with the open source version of the
\verb|C++| library \verb|Qt5|.

To use the software, after observing or discussing with a patient, and according
to what he observes and the answers of the patient, the physician introduces
signs and symptoms in the QAMDiagnos (at least one sign or symptom and a maximum
of six signs and symptoms). The result and a proposal of treatment are given in
the text box ``Treatment proposal`` for the desktop GUI or in the only one text
box for the smart-phone GUI. Although weight and age can be important for
accurate diagnosis, the QAMDiagnos does not use these data because it can occur
that the physician forgets to take them or not. Therefore, in the text box
``Treatment proposal`` the QAMDiagnos shows a treatment proposal which is not
bound to the weight or age. We made the simulations on a \emph{classical 
computer}, so we can obtain each probability (recognition efficiency).

Thus, the physician can also compare the recognition efficiency of the linear
part of the QAMDiagnos for each disease given in the ''Results'' area. So, the
disease with the greatest recognition efficiency can be viewed as the
corresponding disease of the patient. To achieve this goal, that is to simulate
the probabilistic nature of the quantum theory, we use the \verb|qrand()|
function of \verb|Qt5| that allows us to choose arbitrarily one of the diseases
according to its probability. As the linear QAM algorithm is the main algorithm,
the disease with the highest probability is given as result with the associated
treatment proposal. Before that, the non-linear QAM algorithm identifies if a sign or
symptom specific to a particular disease is present. The results of the linear
QAM algorithm and the non-linear QAM algorithm are given together to help the
physician in his decision.

The complementary of the two algorithms is more highlighted in the case of
polyinfection. For example, if common signs and symptoms to malaria, yellow
fever and dengue are inserted and symptom particular to malaria and another to
yellow fever are also inserted, the linear QAM algorithm will identify this
combination as a polyinfection. But, the non-linear QAM algorithm will identify
the signs and symptoms of malaria and yellow fever. This gives to the physician
the possibility to focus only on two diseases (malaria and yellow fever) instead
of the three.

Some snapshots of the QAMDiagnos allowing better understanding are available as 
supplementary material associated to this paper.

\section{Conclusion}\label{sec:concl}

We have presented in this paper the QAMDiagnos, a Quantum Associative Memory
framework that can be helpful to diagnose four tropical diseases (malaria,
typhoid fever, yellow fever and dengue) with several common signs and symptoms.
The framework that has a friendly multi-platform GUI, is a combination of the
improved versions of original quantum linear retrieving algorithm proposed by
Ventura and the non-linear quantum search algorithm of Abrams and Lloyd. While
the phase-inversion introduced in the original linear QAM algorithm increases
the capacity of the memory to make a good diagnosis, the non-linear QAM
algorithm helps to confirm or to correct the diagnosis and to make some
suggestions to the medical staff for the treatment. In addition, the QAMDiagnos
can distinguish a single infection from a co-infection and needs only few
specific signs and symptoms of disease and common signs and symptoms. Therefore,
the QAMDiagnos framework can be a good tool to help inexperienced medical staff
or medical staff without laboratory facilities to rapidly and accurately
diagnose malaria, typhoid fever, yellow fever and dengue, four tropical diseases
that are often confused. Due to its highly automated nature, health centres
personnel can be trained to operate the QAMDiagnos in just one day.

For future works, it is planned to build up a device that can help acquire 
physiological parameters on a patient and transfer them to the QAMDiagnos.

\section*{Acknowledgements}

We thank Pr. Paul WOAFO for his  helpful discussions and remarks. We also thank 
Dr. OKALA, Dr. DOUALLA and Pr. LUMA of the Hopital G\'en\'eral de Douala 
(Cameroon) for their remarks and their help on the classification of signs and 
symptoms. We also thank Moise SOH, the Senior Translator, for 
proofreading 
our work.

 \bibliographystyle{unsrt}

\appendix

\section{Symptoms of each group of diseases}\label{sec:signs}

\begin{table}[H]
\centering
\begin{scriptsize}
\definecolor{tcA}{rgb}{0.627451,0.627451,0.643137}
\begin{tabular}{p{3.5cm}p{4cm}p{3.5cm}p{3.5cm}}\hline
\textbf{Malaria} & \textbf{Typhoid fever} & \textbf{Yellow fever} 
& \textbf{Dengue}\\\hline
Headache	&Headache	&Headache	&Headache\\
Abdominal pain	&Abdominal pain	&Abdominal pain	
&Abdominal pain\\
Fever	&Fever	&Fever	&Fever\\\hline\hline
Myalgia	&	&Myalgie	&Myalgia\\
Renal failure	&	&Renal failure	&
Renal failure\\
Nausea	&	&Nausea	&Nausea\\
Vomiting	&	&Vomiting	&Vomiting\\\hline\hline
Shiver or cold sensation	&	&Shiver or cold sensation	
&\\
Hemorrhage	&	&Hemorrhage	&\\
Iterus	&	&Iterus	&\\
Oliguria	&	&Oliguria	&\\\hline\hline
Anorexia	&Anorexia	&	&\\
Coma	&Coma	&	&\\
Prostration	&Prostration	&	&\\\hline\hline
	&Epistaxis	&Epistaxis	&\\
	&Myocarditis	&Myocarditis	&\\\hline\hline
	&	&Rachiodynia	&Rachiodynia\\
	&	&Heart failure	&Heart failure\\
	&	&Hepatic failure	&Hepatic failure\\
	&	&Conjunctival injection	&
	Conjunctival injection\\
	&	&Somnolence	&Somnolence\\\hline\hline
Anemia	& Asthenia	& Acidosis	& Accumulation of fluid and respiratory 
distress\\
Shock	&Bouveret's ulcer&Bradycardia 
related to temperature	&Generalized adenopathy\\
Generalized or focal convulsion	
&Relative bradycardia	&Infectious shock
&Agitation or Lethargy\\
Multiple convulsion	&Constipation	&Terminal coma	
&Arthralgia\\
Stiffeness	&Yellowy diarrhea	&Excessive dryness
&Asthénie 
prolongée\\
Delirium	&Encephalitis	&Back pain	
&Elevation of the hematocrit and fast drop of 
platelets\\
Respiratory distress	&Lenticular exanthem of 
limbless man	&Limb pain	&Corrected shock\\
Diarrhea	&Rumble in the right 
iliac cavity	&Renal pain	&Uncorrected shock\\
Hemoglobinuria	&Digestive hemorrage or 
digestive perforation	&Epigastric pain	
&Tensor fall\\
Hepatosplenomegaly	&Insomnia	&Fatigue	
&Desquamation of macular eruption\\
Hypoglycemia	&Sabural tongue	&Foul breath	&Retro-orbital pain\\
Malaise	&Abdominal heavines	&Hypotension	
&Encephalopathy\\
Pulmonary edema	&Peritonitis	&Lumbago	
&Macular eruption\\
Palenes	&Dissociate pulse	&Prolongation of the PR and QT intervals on 
electrocardiography	&Maculopapular exanthem\\
Sub-iterus	&Septicemia	&Light proteinuria	
&Hepatomegaly ($>$2cm)\\
Profuse sweat	&Splenomegaly	&Purpura	
&Sudden hypothermia\\
Conscience trouble	&Moderate splenomegaly	&Vomito 
negro	&Leukopenia\\
	&Typhoid state	&	&Obnubilation\\
	&	&	&Vascular purpura\\
	&	&	&Benign cutaneous or 
mucous bleeding\\
	&	&	&Severe cutaneous or mucous bleeding\\
	&	&	&Persitent vomiting\\\hline
\end{tabular}
\end{scriptsize}
\caption{signs and symptoms of the database \cite{epilly,manson,ecn,msf}.}
\label{tab_allSigns}
\end{table}

\end{document}